\documentclass[sigconf]{acmart}
\usepackage{comment}
\usepackage{subcaption}
\usepackage{algorithm}
\usepackage[noend]{algpseudocode}

\AtBeginDocument{%
  \providecommand\BibTeX{{%
    \normalfont B\kern-0.5em{\scshape i\kern-0.25em b}\kern-0.8em\TeX}}}

\begin{document}

\title{Exo-SIR: An Epidemiological Model to Analyze the Impact of Exogenous Infection of COVID-19 in India}








\author{Nirmal Kumar Sivaraman$^1$, Manas Gaur$^2$, Shivansh Baijal$^1$, Ch V Radha Sai Rupesh$^1$}
\author{Sakthi Balan Muthiah$^1$, Amit Sheth$^2$}
\affiliation{
\institution{\textsuperscript{1}Department of Computer Science and Engineering, The LNM Institute of Information Technology, Jaipur, India}
\institution{\textsuperscript{2}AI Institute, University of South Carolina, USA}}
\email{{nirmal.sivaraman,shivansh.baijal.y16,chvradhasairupesh.y17, sakthi.balan}@lnmiit.ac.in} 
\email{mgaur@email.sc.edu, amit@sc.edu}

\begin{abstract}
Epidemiological models are the mathematical models that capture the dynamics of epidemics. The spread of the virus has two routes - exogenous and endogenous. The exogenous spread is from outside the population under study, and endogenous spread is within the population under study. Although some of the models consider the exogenous source of infection, they have not studied the interplay between exogenous and endogenous spreads. In this paper, we introduce a novel model - the Exo-SIR model that captures both the exogenous and endogenous spread of the virus. We analyze to find out the relationship between endogenous and exogenous infections during the Covid19 pandemic. First, we simulate the Exo-SIR model without assuming any contact network for the population. Second, simulate it by assuming that the contact network is a scale free network. Third, we implemented the Exo-SIR model on a real dataset regarding Covid19. We found that endogenous infection is influenced by even a minimal rate of exogenous infection. Also, we found that in the presence of exogenous infection, the endogenous infection peak becomes higher, and the peak occurs earlier. This means that if we consider our response to a pandemic like Covid19, we should be prepared for an earlier and higher number of cases than the SIR model suggests if there are the exogenous source(s) of infection. 
\end{abstract}




\maketitle

\section{Introduction}
The \emph{Susceptible, Infected and Recovered} model (SIR model) is considered as one of the seminal models of pandemics \cite{guille2013information}. The model describes how the spread of the virus from people to people within the population under consideration. However, the spread of infection is not only from within the community but also from the external sources. According to WHO, the causes of infection within the population and external to the population are called as \emph{Local transmission} and \emph{Imported cases}, respectively. We call them as an endogenous and exogenous spread of infection, respectively.


If we consider the Covid19 pandemic in India as an example, a major sub-event of Covid19 spread happened -- the Tablighi Jamaat religious congregation. It was held in Delhi from $1^{st}$ March to $21^{st}$ March $2020$\footnote{\url{https://en.wikipedia.org/wiki/2020_Tablighi_Jamaat_coronavirus_hotspot_in_Delhi}} despite the Government of India imposed nationwide restrictions on holding a meeting with a large number of people. Over $9000$ people participated in that event\footnote{ "Coronavirus: About $9000$ Tablighi Jamaat members, primary contacts quarantined in-country, MHA says". The Times of India. Press Trust of India. $2^{nd}$ April $2020$.}. $4291$  cases have been reported that can be traced to the event\footnote{"Coronavirus | Nearly $4300$ cases were linked to Tablighi Jamaat event, says Health Ministry". The Hindu. Press Trust of India. $18^{th}$ April $2020$.}. As of $18^{th}$ April $2020$, $30\%$ of the cases in India were due to this event \footnote{ABP News Bureau (18 April 2020). "Tablighi Jamaat Responsible For $30\%$ Total Coronavirus Cases in India: Health Ministry". ABP News.}. The mobility of people who attended this event to their states during Covid19 has been causative phenomena that propelled the spread of the virus. 

Another major sub-event is the mass movement of migrated workers across the country, as we are writing this paper\footnote{https://en.wikipedia.org/wiki/Indian\_migrant\_workers\_during\_the\_COVID-19\_pandemic}. These migrant workers who are returning to their home states are the exogenous source of infection for the population in those states.

Hence, while studying, modeling the exogenous sources of infection is essential. As can be seen from the examples discussed here, not modeling the external sources of contamination could give rise to an inaccurate estimation of the scenario. However, in the SIR model \cite{kermack1927contribution, hethcote1989three}, sources of infection exogenous to the population under study are not considered. These models assume that there are a finite number of infections in the closed community itself at the initial stage. With that assumption, it predicts the number of people infected at a later instant of time. 

Another issue with the SIR model is that it assumes that in the initial state, there is a set of nodes that have already been infected. Also, the SIR model is extremely dependent on the initial state. It considers the entire population to be susceptible except a few that are infected. It then predicts the spread. If there are no infected people in the initial state, then the model will predict that no one will ever be infected.

\begin{figure*}[ht]
 	
 	\includegraphics[width=.65\textwidth]{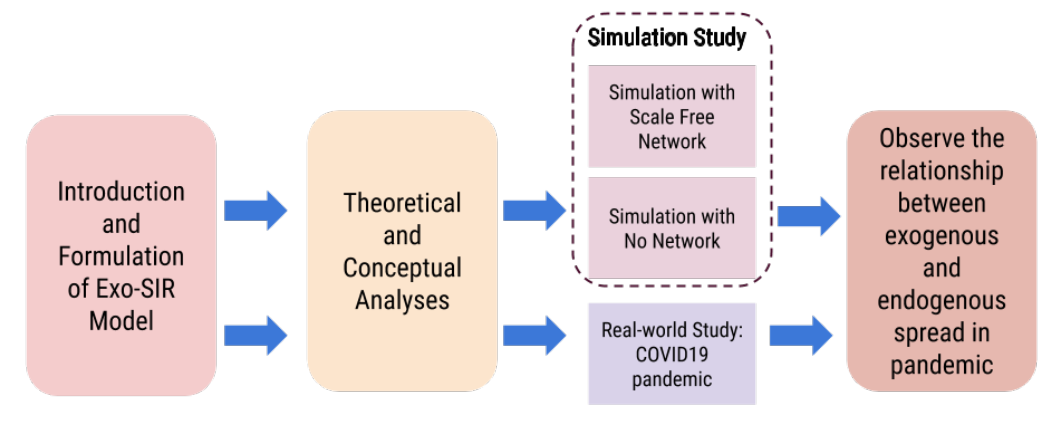}
 	\caption{Summary of our approach to study the relationship between exogenous and endogenous spread.}
 	\label{summary}
 \end{figure*}

Addressing these concerns, we propose the Exo-SIR model that extends the SIR model to consider the endogenous and exogenous spread. We simulate the model in two ways; 1) assuming that the contact network of people is a scale free network, 2) assuming no network (a well-mixed population). We simulate the coefficients' different values to figure out the relationship between the exogenous and endogenous spreads. We also apply the Exo-SIR model on a real dataset regarding the spread of the Covid19 pandemic in the Indian states of Rajasthan, Tamil Nadu, and Kerala. Exogenous spread dominates endogenous spread in Tamil Nadu, whereas the contrary is true in the case of Rajasthan. In Kerala, both the endogenous and exogenous spread have roughly the same prevalence. The trends in the theoretical analysis, results of the simulations, and the analysis of the real dataset are consistent. The summary of our approach is given in Figure~\ref{summary}. The details of the dataset that we used to infer these relations are given in Section~\ref{data}.

This paper is structured as follows. The next two sections, discuss related works and preliminaries. In Section~\ref{the_model}, we describe the Exo-SIR model. Next we compare the Exo-SIR model with the SIR model. In Section~\ref{endo-exo}, we discuss the relationship between endogenous and exogenous spreads. In the section after that, we explain the two types of simulations and explain the results. In Section~\ref{simulation}, we discuss the analysis on real data. 

\section{Related Works}
\label{Re}
Here, we discuss the works that are related to the idea of exogenous influence to the population under study.

\subsection{Epidemiological Models}

Two popular epidemiological models are SEIR model \cite{biswas2014seir} and SEYAR model \cite{aguilar2020investigating}. In case of SEIR model, users are classified into four states. In SEIR model, users are classified into four -- Susceptible (who are prone to infection), Exposed (who are infected, but not infectious), Infected and Recovered. 

In case of SEYAR model, there are more states compared to SEIR model. In SEYAR model, users are classified into five -- Susceptible (who are prone to infection), Exposed (who are infected, but not infectious), Infected and symptomatic (shows the symptoms), Infected asymptomatic (shows no symptoms) and Recovered.

Even though the SEIR model and SEYAR model are popular models, they do not consider any external source of infection. Also, the state Exposed (who are infected, but not infectious) is not appropriate in case of Covid19 as there is no latency \cite{hamzah2020coronatracker}. 

Also, we can see in \cite{peiliang2020seir,hamzah2020coronatracker} the following:  

\begin{quote}
In a small number of case reports and studies, pre-symptomatic transmission  has been documented through contact tracing efforts and enhanced investigation of clusters of confirmed cases. This is supported by data suggesting that some people can test positive  for COVID-19 from 1-3 days before they develop symptoms. Thus, it is possible  that people  infected  with  COVID-19 could  transmit the  virus  before  significant  symptoms  develop. 
\end{quote}

Even though we cannot rule out the possibility of pre-symptomatic and asymptomatic transmission, we can see from the quoted text that they are very small. Hence, we assume that pre-symptomatic and asymptomatic transmission are negligible in case of Covid 19.

\subsection{Models of External Influence on online social networks}
Information diffusion in online social networks is similar to the way the virus spreads in a population. There are a few recent works in the literature that attempt to model the external influence in information diffusion in online social networks \cite{myers2012information}. \cite{myers2012information} and \cite{li2015measuring} propose information diffusion model on the network. In these works, they assume that the information flows through an underlining network. Also, they consider links from other web sites as the external sources of information. The internal diffusion are those when the messages that are shared do not have any external links. 

The work described in \cite{myers2012information} uses very specific parameters like the following:
\begin{itemize}
\item probability of any node receiving an exposure at time t
\item the random amount of time it takes an infected node to expose its neighbors
\item how the probability of infection changes with each exposure
\item the probability that the node i has received n exposures by time t
\end{itemize}

Whereas, the work described in \cite{li2015measuring} trace the cascade and reconstruct the graph as much as possible. Also, they conclude that external influence has bigger impact on the network when compared to the influence of the social media influencers.

The model that is closest to our work is Yang et al's model \cite{yang2019modeling, Shayak:KiML2020}, which is an extension of the SIR model (explained in Section \ref{prelim}) to include the external influence to the network. 

\begin{figure}[h]
	\includegraphics[width=.48\textwidth]{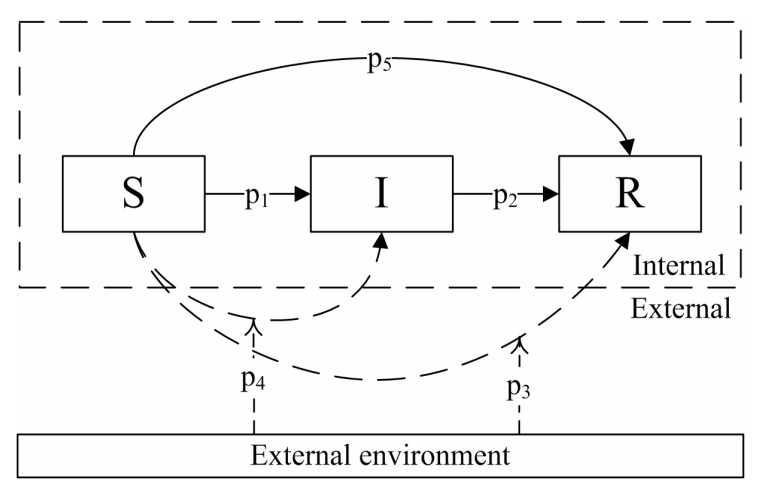}
	\caption{State transition diagram of the diffusion mechanisms in Yang et al's model. Diagram taken from \cite{yang2019modeling}.}
	\label{f1}
\end{figure}


\begin{equation}\label{key01}
s+i+r = 1
\end{equation}

\begin{equation}\label{key02}
\frac{ds}{dt} = -p_{1}ksi-((1-p_1)p_3+p_4)\theta s-(1-p_1)p_5ksi
\end{equation}

\begin{equation}\label{key03}
\frac{di}{dt} = p_{1}ksi+p_4\theta s-p_2i
\end{equation}

\begin{equation}\label{key04}
\frac{dr}{dt} = p_2i+(1-p_4)p_3\theta s+(1-p_1)p_5ksi
\end{equation}

There are two possible transitions from the state S to I. One path is the normal endogenous path, and the second one is due to external influence. These transitions have probabilities $p_1$ and $p4$, respectively. Similarly, there are two possible transitions from the state S to R -- one through endogenous and the other through external influence. Their probabilities are $p_5$ and $p_3$, respectively. However, the transition from state I to R does not affect external influence(s).

Although the exogenous infection is modeled in Yang et al.'s model, it fails to capture the dynamics between the endogenous and exogenous infections.

\subsection{Other Studies of Endogenous and Exogenous Information Diffusion}

The dual nature of message flow over the online social network is studied and verified in \cite{de2018demarcating}. Here, the dual nature refers to the injection of exogenous opinions to the network and the endogenous influence-based dynamics\cite{sivaraman2017social}. In \cite{fujita2018identifying}, the authors propose a method for extracting the relative contributions of exogenous and endogenous contents. In \cite{agrawal2012learning}, the authors postulate that the nature of the information plays a crucial role in the way it spreads through the network. They quantify two properties of the information -- endogeneity and exogeneity. Endogeneity refers to its tendency to spread primarily through the connections between nodes and exogeneity refers to its tendency to spread to the nodes, independently of the underlying network. In \cite{oka2014self}, the authors study the bursts that originate from endogenous and exogenous sources and their temporal relationship with baseline fluctuations in the volume of tweets. The study reported in \cite{crane2008robust} classifies the bursts into endogenous and exogenous. According to this study, those bursts that reach the peak almost instantaneously after the diffusion starts and then go down slowly are exogenous burst. Also, those bursts that gradually increase and slowly decrease are endogenous. 

 	
\section{Preliminaries}
\label{prelim}
In this section, we briefly review the SIR epidemiological model. It describes how epidemics spread through a population. This model is used for the study of information diffusion approximating that the way epidemics spread and the way information gets diffused in a population are similar. 







In this model, the population is classified into three -- Susceptible (who are prone to infection), Infected (who contain the infection), and Recovered (who do not have the infection and its associated symptoms). In the limit of sizeable total population $N$ that does not change over time, the given equations models the dynamics of the spread \cite{bailey1975mathematical}: 

\begin{equation}\label{sir1}
s(t)+i(t)+r(t) = 1
\end{equation}

\begin{equation}\label{sir2}
\frac{ds}{dt} = -\beta si
\end{equation}

\begin{equation}\label{sir3}
\frac{di}{dt} = \beta si - \gamma i
\end{equation}

\begin{equation}\label{sir4}
\frac{dr}{dt} = \gamma i
\end{equation}

where the fraction of Susceptible, Infected and Recovered people at time $t$ are represented by $s(t),  i(t)$ and $r(t)$ respectively. $\beta$ is the rate of infection, and $\gamma$ is the rate of recovery.

\section{The Model}
\label{the_model}
The Exo-SIR model is an extension of SIR model in which the nodes have Susceptible, Infected and Recovered states. It differs from SIR model in the following ways:
\begin{itemize}
	\item It classifies infected nodes into two different types -- Infected from exogenous source and Infected from endogenous source.
	\item It differentiates between the spread from endogenous and exogenous sources. 
\end{itemize}

Susceptible nodes become infected with a certain probability called the rate of infection. This rate could be different for endogenous and exogenous infections. The nodes that got affected by endogenous source and exogenous sources move into different states. We assume that susceptible nodes get infected from only one of these sources and never from both the sources. Hence, even when some nodes are susceptible to both endogenous and exogenous infection, they become infected by either an endogenous or an exogenous source. The infected nodes recover with a certain probability called the recovery rate. These nodes move into the recovered state. The advantage of the Exo-SIR model compared to the SIR model is that we can observe the endogenous and exogenous diffusion separately.

We use the following notations:
\begin{itemize}
    
	\item[$S$] state of susceptible
	\item[$I_x$] state of infected from exogenous source
	\item[$I_e$] state of infected from endogenous source
	\item[$R$] state of recovered
	\item[$i_x$] Fraction of nodes that are infected from exogenous source
	\item[$i_e$] Fraction of nodes that are infected from endogenous source
	\item[$r$] Fraction of nodes that are recovered
	\item[$\beta_x$] Rate at which the exogenous source infects the nodes
	\item[$\beta_e$] Rate at which the nodes infects other nodes
	\item[$\gamma$] Rate at which the nodes get recovered
	\item[] We use the words infection, diffusion and spread interchangeably according to the context.
\end{itemize}

\begin{figure}[h]
	\centering
	\includegraphics[width=.20\textwidth]{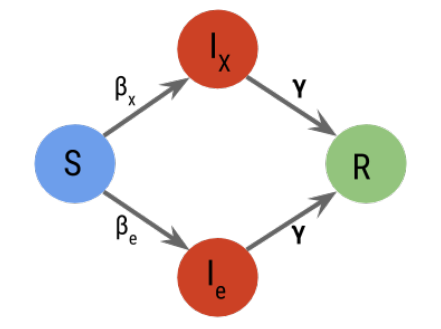}
	\caption{State transition diagram of the nodes in the Exo-SIR model}
	\label{f4}
\end{figure}

The state transition diagram of the Exo-SIR model is given in Figure~\ref{f4}.

We classify infected nodes into two different types -- infected from exogenous source and infected from endogenous source.
\begin{equation}\label{key2}
  i_e+i_x = i
\end{equation}

We assume that the total population remains constant.
\begin{equation}\label{key3}
  s+i+r = 1
\end{equation}


a fraction of the susceptible people ($s$) gets infected by exogenous source and another fraction of $s$ gets infected by endogenous sources. For endogenous infection, the population that is infected plays a big role. Hence, we can say that:
\begin{equation}\label{key4}
  \frac{ds}{dt} =  -\beta_x s -\beta_e s i
\end{equation}

Change in $i_{x}$ is influenced by all the susceptible nodes that are prone to the exogenous influence.

\begin{equation}\label{key9}
  \frac{di_x}{dt} = \beta_x s - \gamma i_x
\end{equation}

Change in $i_{e}$ is influenced by all the susceptible nodes and infected nodes. 

\begin{equation}\label{key10}
  \frac{di_e}{dt} = \beta_e s i - \gamma i_e
\end{equation}

Change in $r$ depends on the number of infected people in the network.
\begin{equation}\label{11}
  \frac{dr}{dt} = \gamma i\end{equation}

\section{Comparison with SIR model}

Mirroring the rate of change of s(t), i(t), and r(t) in the SIR model (Section~\ref{Re}), we find the expressions for the rate of change of s(t), i(t), and r(t) for the Exo-SIR model. 


\subsubsection{Rate of Change of s:}

\begin{equation}\label{key8}
  \frac{ds}{dt} = -\beta_x s-\beta_e s i
\end{equation}

\subsubsection{Rate of Change of r:}

\begin{equation}\label{17}
\frac{dr}{dt} = \gamma i
\end{equation}

\subsubsection{Rate of Change of i:}

Differentiating Equation~\ref{key2} with respect to time, we get

\begin{equation}\label{key11}
\frac{di}{dt} = \frac{di_e}{dt}+\frac{di_x}{dt}
\end{equation}

\begin{equation}\label{key12}
  \frac{di}{dt} = \beta_e s i - \gamma i_e + \beta_x s - \gamma i_x
\end{equation}

\begin{equation}\label{key14}
  \frac{di}{dt} = \beta_e s i + \beta_x s - \gamma (i_x + i_e)
\end{equation}

Applying Equation~\ref{key2} on Equation~\ref{key14}, we get
\begin{equation}\label{key15}
\frac{di}{dt} = \beta_e s (i_x + i_e) + \beta_x s - \gamma (i_x + i_e)
\end{equation}


Here, even if we assume that there is no infected people in the beginning -- i.e. $i_e = 0$ and $i_x = 0$, we get the following.

Applying Equation~\ref{key2} on Equation~\ref{key14}, we get
\begin{equation}\label{key16}
\frac{di}{dt} =  \beta_x s
\end{equation}

This shows that unlike the SIR model, the Exo-SIR model models the situation if no one is infected initially. SIR model assumes that there is an initial outbreak size  $i_0$. This means $i_0$ people are infected in the beginning and $i_0 > 0$ \cite{hethcote2000mathematics}. Our work addresses this limitation of the SIR model. Note that, the Exo-SIR model would behave the same way as the SIR model if we assume that $i_x = 0\ and\ \beta_x = 0$.

\section{Dynamics of Exogenous Spread and Endogenous Spread}
\label{endo-exo}
In this section, we find the relationship between the cumulative exogenous infections ($i_x$) and the daily endogenous infections ($\frac{di_e}{dt}$).

Applying Equation~\ref{key2} on Equation~\ref{key10}, we get

\begin{equation}\label{18}
\frac{di_e}{dt} = \beta_e s(i_e+i_x) - \gamma i_e
\end{equation}

\begin{equation}\label{13}
\frac{di_e}{dt}\bigg\rvert_{i_x>0} = \beta_e s(i_e+i_x) - \gamma i_e
\end{equation}

At $i_x = 0$, 
\begin{equation}\label{14}
\frac{di_e}{dt}\bigg\rvert_{i_x=0} = \beta_e si_e - \gamma i_e
\end{equation}

Since all $\beta_e, s, i_e, and \gamma$ are positive, 
\begin{equation}\label{15}
\frac{di_e}{dt}\bigg\rvert_{i_x=0} < \ \frac{di_e}{dt}\bigg\rvert_{i_x>0}
\end{equation}

This shows that $\frac{di_e}{dt}$ increases in the presence of $i_x$. In other words, this shows that the presence of exogenous diffusion causes endogenous diffusion to increase. In the next section, we discuss simulations that we did to verify this.

\section{Simulation}
\label{simulation}
We simulate the Exo-SIR model to determine its behavior for a variety of scenarios that are represented by the different values of its parameters. We simulated the model in two ways: 

\begin{enumerate}
    \item By assuming that the people network is scale free network. Within this network, the susceptible nodes can catch the infection from only those infected nodes, which they are connected to through an edge, i.e., their immediate neighbors 
    \item By assuming no network (well-mixed population). Contrary to the above, in this scenario, a susceptible node can get infected from any of the infected nodes in the population under consideration.
\end{enumerate}

We chose Barab{\'a}si-Albert network because there are pieces of evidence that the human disease network could be scale free \cite{szabo2020propagation}. The results of these are discussed in the following section.

\subsection{Using Scale free Network}
\begin{figure*}
\begin{subfigure}{.35\textwidth}
  \centering
   	\includegraphics[height=1.5in,width=.9\textwidth]{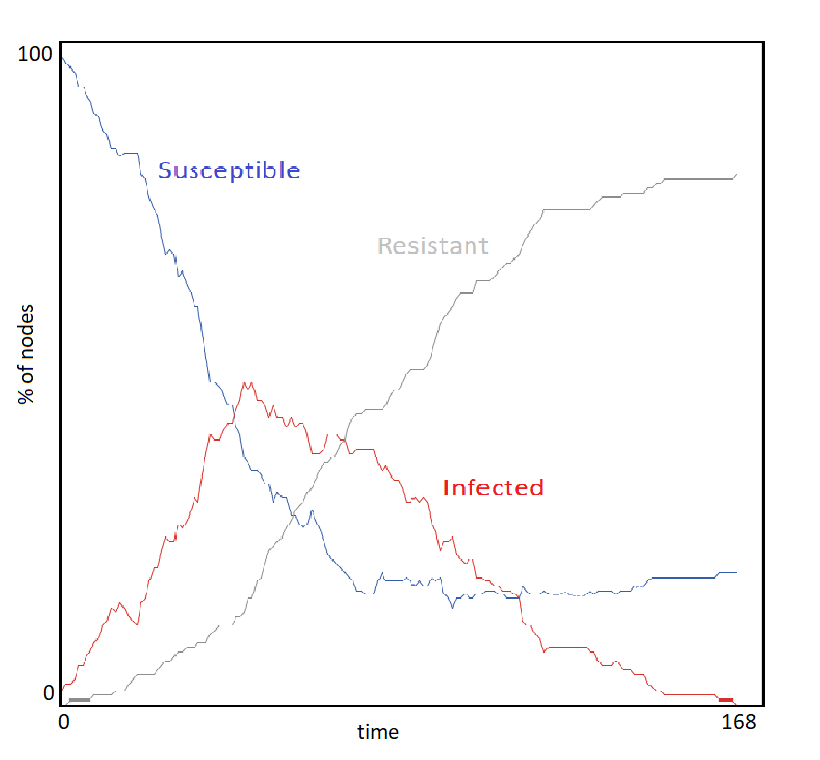}
 	\caption{plot of s, i and r with no exogenous source}
 	\label{f5}
\end{subfigure}%
\begin{subfigure}{.35\textwidth}
  \centering
	\includegraphics[height=1.5in,width=.9\textwidth]{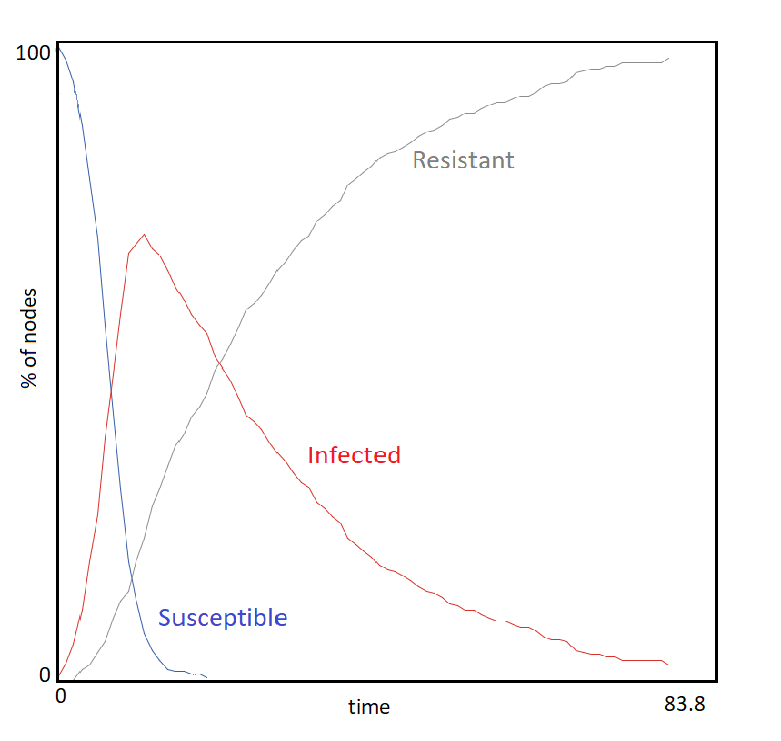}
 	\caption{plot of s, i and r with exogenous source}
 	\label{f6}
\end{subfigure}
\caption{Plots of the simulation. In the presence of exogenous source, the peak is becoming more significant and occurring early. Accordingly, the humanitarian authorities can leverage the Exo-SIR model to estimate the peak more precisely than the SIR model. While ignoring exogenous factors due to the adoption of the traditional SIR model, the authorities would be caught off guard as they will be preparing for a later and smaller peak.}
\label{fig:fig1}
\end{figure*}

The analysis presented in this section has been done considering a scale free contact network for the population under study, which is called Barab{\'a}si-Albert network \cite{barabasi2013network}. Under this scenario, the susceptible nodes can catch the infection from only those infected nodes, which they are connected to through an edge, i.e., their immediate neighbors. We have predicted the values for various combinations of $\beta_x$, $\beta_e$, and $\gamma$ using the Exo-SIR model in the network mentioned above. We used NetLogo Simulation tool \cite{tisue2004netlogo} to carry out the predictions. Sample simulation results are shown in Figures~\ref{f5} and ~\ref{f6}. Figure~\ref{f5} shows the SIR model's simulation results with no exogenous influence, and Figure~\ref{f6} shows the simulation results with exogenous influence.

Here we can see that, when we consider exogenous factors, the peak of the distribution of the number of the infected population shows significant changes in the following ways:

\begin{itemize}
    \item the height of the peak (peak value) increases
    \item the peak (peak tick) occurs early
\end{itemize}



\subsubsection{Dynamics of Exogenous Spread on Endogenous Spread}
Next, we would like to determine the changes in endogenous spread with the change in exogenous factors.
The step by step methodology adopted to carry out the simulation, and the analysis is given in Algorithm~\ref{a3}.

\begin{algorithm}[h]
    
    \begin{algorithmic}[1]
        
        \State Initialize $\beta_x$, $\beta_e$ and $\gamma$ with 3 different values, i.e., 0.1, 0.5 and 0.9. Henceforth, we have 27 different combination of these parameters.
        \State For each of the combinations of $\beta_x$, $\beta_e$, and $\gamma$, iterate over steps 3 and 4 fifty times.
        \State Setup a Barab{\'a}si-Albert network of 150 nodes having an average node degree of 2 \cite{barabasi2013network}.
        \State Simulate and predict the values of S, $I_e$, $I_x$, and R using the Exo-SIR model. 
        \State Extract the values of endogenous peak and its time slice, exogenous peak, and time slice from each of the simulations.
        \State Calculate the mean peak value and peak tick of Exo and Endo nodes so that we have one value per combination of $\beta_x$, $\beta_e$, and $\gamma$.
    \end{algorithmic}
\caption{Algorithm to perform the simulations and analysis by assuming that the contact network in the population is scale-free }\label{a3}
\end{algorithm}

In the above algorithm, we have carried out 50 simulations for each combination of the parameters and averaged it out to address the bias that might get introduced due to the structure of the network since the setting up of a network in step 3 in the above algorithm is random each time.

\begin{figure*}
\begin{subfigure}{.35\textwidth}
  \centering
   	\includegraphics[height=1.5in,width=.9\textwidth]{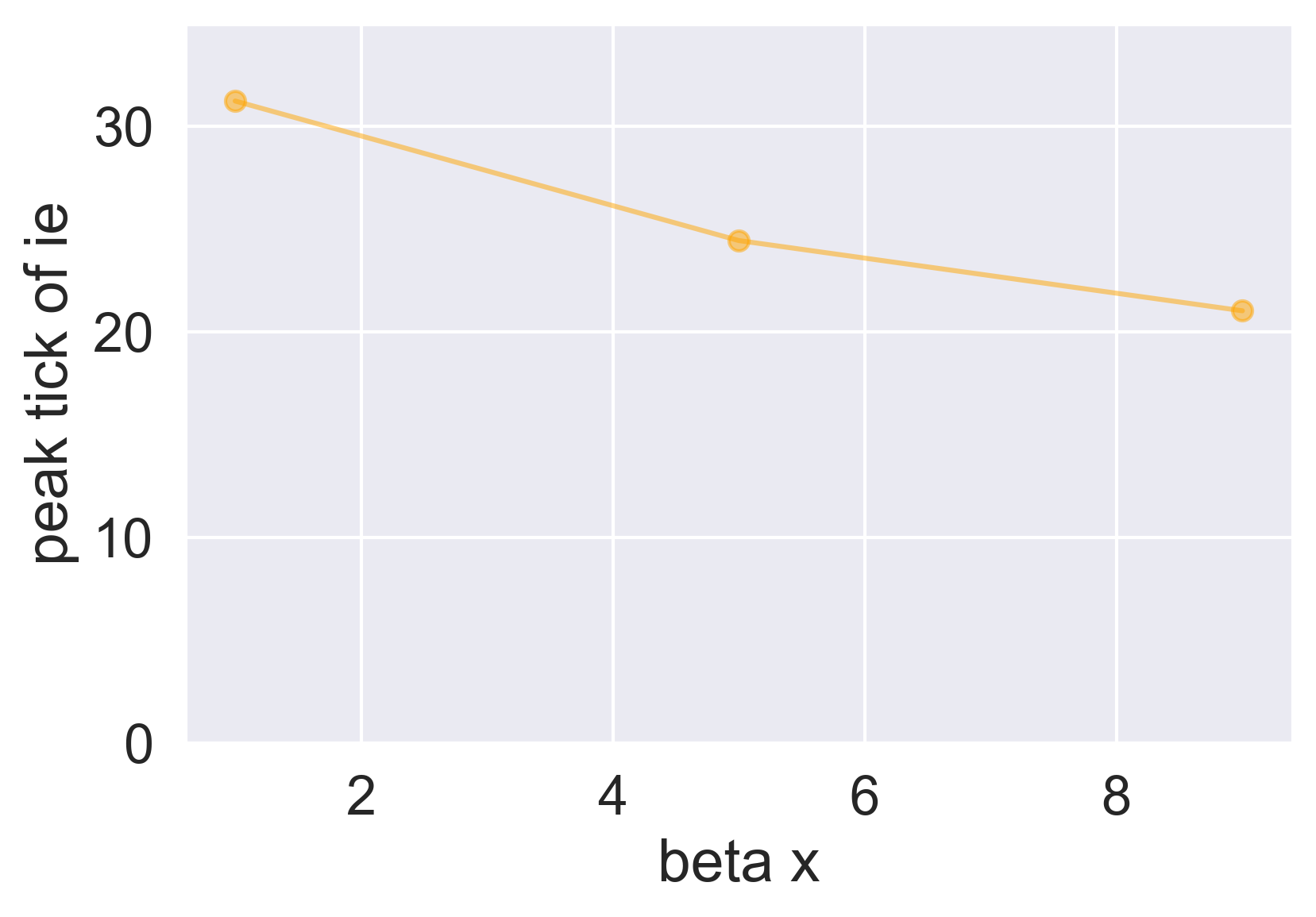}
 	\caption{impact of $\beta_x$ on peak tick}
 	\label{f200}
\end{subfigure}%
\begin{subfigure}{.35\textwidth}
  \centering
	\includegraphics[height=1.5in,width=.9\textwidth]{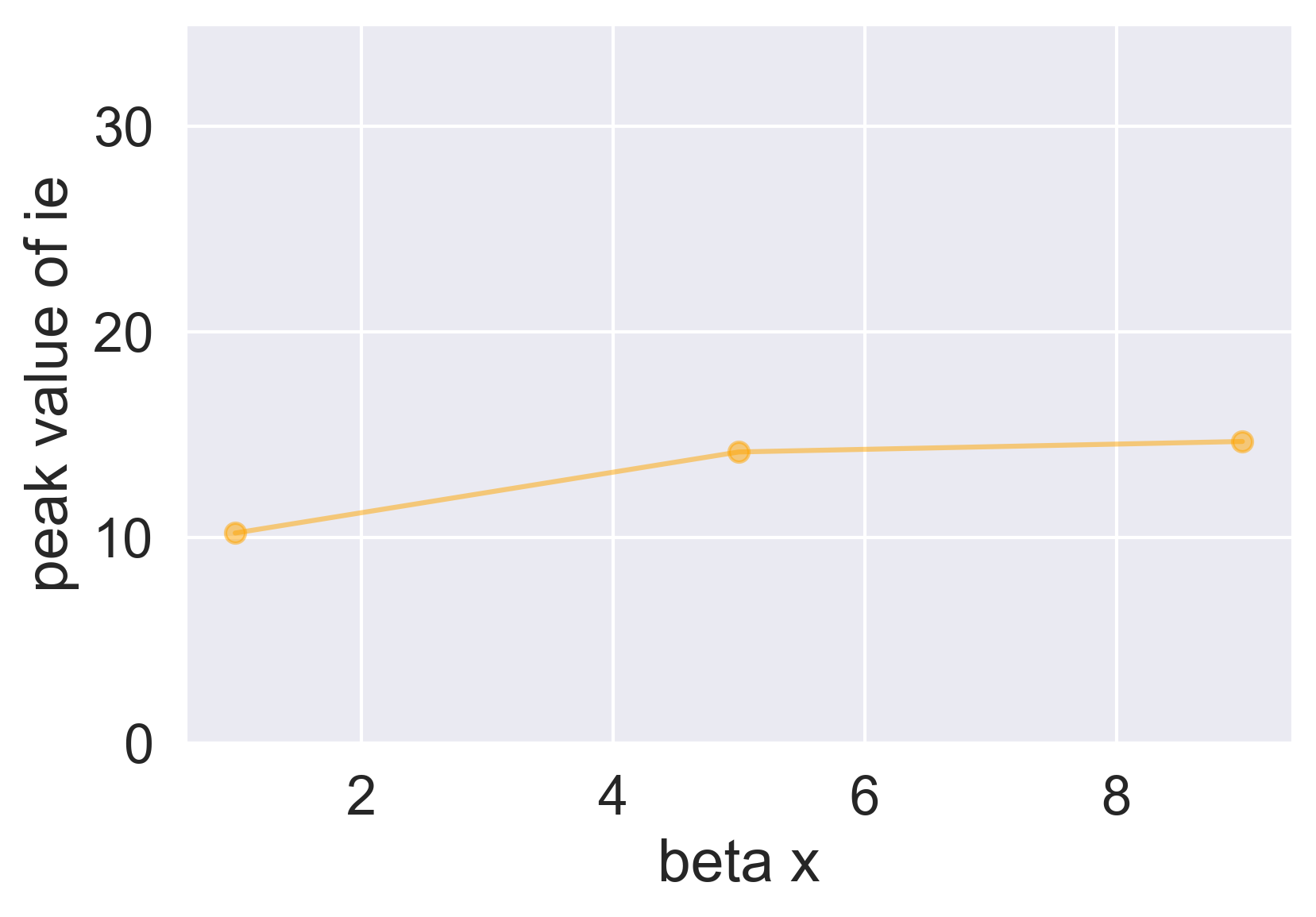}
 	\caption{impact of $\beta_x$ on peak value}
 	\label{f201}
\end{subfigure}
\caption{Plots of the impact of $\beta_x$ on $i_e$. Each dot represents the average of the values obtained from $50$ iterations}
\label{fig:fig2}
\end{figure*}

Figures~\ref{f200} and ~\ref{f201} are a result of simulation and analysis done as described in Algorithm~\ref{a3} and provide us with the following insights:

\begin{itemize}
	
	\item \noindent \emph{Peak Value:} Figure ~\ref{f200} shows that $\beta_x$(exogenous factors) influence the peak value of endogenous infections. The endogenous peak value increases with increase in $\beta_x$
	\item \noindent \emph{Peak tick:} On the other hand, Figure ~\ref{f201} shows that endogenous peak tick decreases with increase in $\beta_x$ 
\end{itemize}

We can conclude that exogenous source and its infection impacts the spread in the network by advancing the peak and increasing the height of the endogenous peak.



\begin{table}
  \caption{Impact of $\beta_e$, $\beta_x$ and $\gamma$ on $ln$(\texttt{ie\char`_peak})}
  \label{t0}
  \begin{tabular}{|l|l|l|l|}
    \hline
      & coef & std err & confidence interval \\
    \hline
    $\beta_e$ & $0.6319$ & $0.006$ & $0.6204 $ to $0.6434$ \\
    $\beta_x$ & $0.6319$ & $0.006$ & $0.6204 $ to $0.6434$ \\
    $\gamma$ & $-0.4390$ & $0.006$ & $-0.4505 $ to $-0.4274$ \\
    \hline
  \end{tabular}
\end{table}

\subsection{With No Network}
In this section, we determine the relative effects of $\beta_x$, $\beta_e$ and $\gamma$ on the endogenous peak statistically and measure the impact of $\beta_x$ on endogenous infections, which is consistent with the results shown above. Here, we did not assume any network for our population, and the objective of these simulations was to determine the impact of $\beta_x$, $\beta_e$ and $\gamma$ on endogenous peak value and peak tick (see Table \ref{t0}). To achieve this, we took a sample of 27000 simulations and analyzed them as described in Algorithm~\ref{a2}.

\begin{algorithm}[h]
    
    \begin{algorithmic}[1]
        
        \State Initialize $\beta_x$, $\beta_e$ and $\gamma$ with 30 random values between 0 and 1 uniformly. 
        \State Initialize the initial number of susceptible, infected(endo and exo) and recovered nodes experimentally as: N = 1000000.0, $S_0$ = 999996.0, $I_{x_0}$ = 3.0, $I_{e_0}$ = 1.0 and $R_0$ = 0.0.
        \State For each of the 27000 combinations of $\beta_x$, $\beta_e$, and $\gamma$, with the above initial condition, we predicted the endogenous and exogenous peak value and peak tick using the Exo-SIR model.
        \State Then, we computed the natural logarithm of the peak value and scaled it between 0 and 1.
        \State Finally, we fitted an OLS Regression Model with $\beta_x$, $\beta_e$ and $\gamma$ as the independent variable and $\ln$(\texttt{Ie\char`_peak}) and the independent variable and analysed the coefficients statistically.
    \end{algorithmic}
\caption{Algorithm to perform the simulations and analysis by assuming no contact network in the population. \\\textbf{Note:} If we look at the differential equations, the system is not a linear one, but rather exponential. Therefore, we took $\ln$(\texttt{Ie\char`\_peak}) as the dependent variable. Table 1 shows the impact of the above three independent variables on the dependent variable.}\label{a2}
\end{algorithm}
 
The following inferences can be drawn from the results of Regression Analysis.
\begin{enumerate}
    \item The p-value for all the three variables is less than 0.05. This means we would reject the null hypothesis, and adopt the alternate hypothesis that the impact of all the three parameters on the peak endogenous infection's peak is statistically significant.
    \item The adjusted R-squared value is maximum(0.70) when all the three parameters are considered while fitting the regression model. This means that we can better explain the variation in the dependent variable when considering all three, i.e., $\beta_e$, $\beta_x$ and $\gamma$. Removing any one of them would decrease the adjusted R-squared value. Also, the confidence interval of each parameter is mentioned in Table 1.
    \item $\beta_x$ impacts endogenous infections as much as $\beta_e$(the contribution of both is almost equal), which is an important observation. This means that exogenous factors also have a considerable impact on the endogenous infection and ignoring the exogenous factors would not give an accurate estimate of the endogenous infections. 
\end{enumerate}

\section{Analysis Using Real Data}

In this section, we describe the data and the analysis of the implementation of SIR model and Exo-SIR model on the Covid19 pandemic. We analyzed the data of three states in India -- Tamil Nadu, Rajasthan and Kerala. The reason for choosing these states is that $i_e \ll i_x$ in Tamil Nadu, $i_e \gg i_x$ in Rajasthan and $i_e \approx i_x$ in Kerala.

\subsection{Datasets}
\label{data}

\begin{figure}[h]
	\centering
	\includegraphics[width=.48\textwidth]{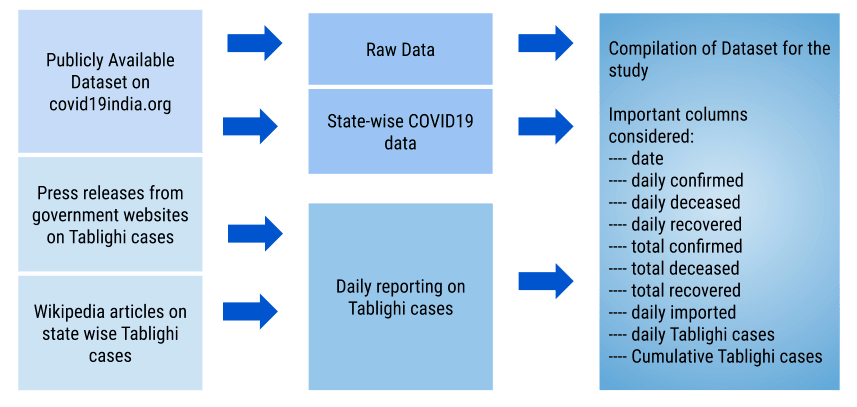}
	\caption{The data pipeline}
	\label{pipeline}
\end{figure}

The data pipeline is given in Figure~\ref{pipeline}. We constructed our dataset from three different sources\footnote{The code and all the data used in our experiments are publicly available at \url{https://github.com/baijalshivansh/Exo-SIR-Model}.} for our analysis -- \url{covid19india.org}, the Government web site of Tamil Nadu for their press release to find the daily number of Tablighi cases and Wikipedia page on state wise daily data. 

\url{covid19india.org} is a publicly available volunteer-driven dataset of Covid19 statistics in India\footnote{\url{https://covid19india.org/}}. There are multiple files in this dataset. One of which is called \emph{raw data} that captures the anonymised details of the patients. In the raw data, the columns of interest for us are DateAnnounced, DetectedState and TypeOfTransmission.


Another file from same source is called states\_daily. In this file the columns of interest are states\_daily/status, states\_daily/kl, states\_daily/rj, states\_daily/tn and states\_daily/date. kl, rj and tn are the codes used in this dataset for the states of Kerala, Rajasthan and Tamil Nadu respectively.

Here, status can have the following values: infected, recovered and diseased. From these columns, we prepared the time series dataset for each state. The columns available in the dataset that we created are daily confirmed, daily deceased, daily recovered, date, total confirmed, total deceased, total recovered and daily imported cases.


Another dataset that we used is the compilation of the press releases (news bulletins) from the Governments of the states under study. This is to get the daily number of cases due to a major sub event of Covid19 spread in Inidia -- Tablighi Jamaat religious congregation. We manually went through the press releases and collected the data.



\subsection{Mapping}
In this section, we discuss how the values in the dataset is mapped on to the variables in the Exo-SIR model. On a particular day, say day $k$,

\begin{equation}
s(t)+i(t)+r(t) = 1
\end{equation}

\begin{equation}
\frac{ds}{dt} = -(\frac{di}{dt} + \frac{dr}{dt})
\end{equation}
\noindent $\frac{di_e}{dt} =$ daily confirmed cases on day $k$

\noindent $\frac{di_x}{dt} =$ daily imported cases on day $k$ + daily cases due to Tablighi event on day $k$ .%

\noindent $\frac{di}{dt} = $ $\frac{di_e}{dt} + \frac{di_x}{dt}$ .

\noindent $\frac{dr}{dt} = $(daily recovered + daily deceased) on day $k$ 

\begin{equation}\label{sir21}
s = 1 - \frac{daily\ confirmed\ on\ day\ 0}{N} 
\end{equation}
where N is the total population who are prone to the infection. 

\begin{equation}\label{sir22}
i = total\ confirmed
\end{equation}
\begin{equation}\label{sir23}
r = total\ deceased\ +\ total\ recovered
\end{equation}

\begin{algorithm}[h]
	
	\begin{algorithmic}[1]
		
		\State For each time slice, we calculate the values of $ \frac{di}{dt}$ and $\frac{dr}{dt}$ from the dataset.
        \State We consider $s$ as the susceptible people from the population of the state under study. 
        \State We calculate the cumulative values $i$ and $r$. 
        \State We find $\gamma$, $\beta_e$ and $\beta_x$ using values of the time for which the data is available.
        \State We ran the Exo-SIR model with these values as the initial values and plotted $i_e$ in the presence of $i_x$ and $i_e$ in the absence of $i_x$
	\end{algorithmic}
\caption{Algorithm to plot the Exo-SIR model.}\label{a1} 
\end{algorithm} 

\subsection{Applying Exo-SIR model on real data}

In this section, we analyze the data from the states of Tamil Nadu, Rajasthan, and Kerala. We compare the peak tick and peak value of the plot of $i_e$ in the presence and absence of $i_x$. This would give information about the impact of $i_x$ on $i_e$. For this purpose, we used Algorithm~\ref{a1}.

\begin{figure*}
\begin{subfigure}{.33\textwidth}
  \centering
   	\includegraphics[width=\textwidth]{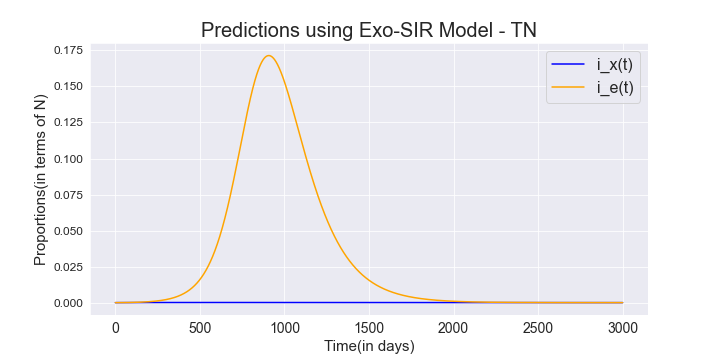}
 	\caption{Exo-SIR model with $i_x$. The values of $i_x$ are very small for the scale of this plot. Hence it is plotted separately. Please refer the Figure~\ref{f108}}
 	\label{f110}
\end{subfigure}%
\begin{subfigure}{.33\textwidth}
  \centering
	\includegraphics[width=\textwidth]{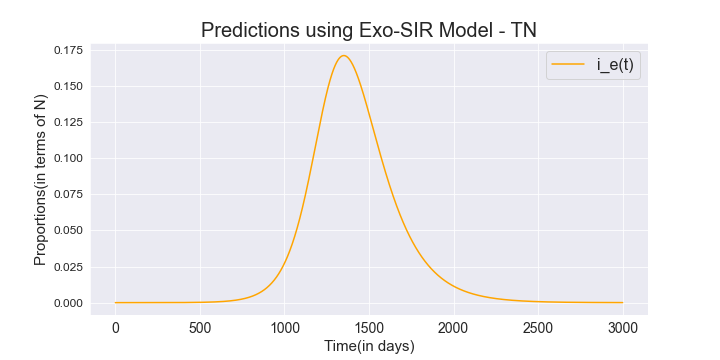}
 	\caption{Exo-SIR model without $i_x$ }
 	\label{f109}
\end{subfigure}
\begin{subfigure}{.33\textwidth}
  \centering
  \includegraphics[width=\textwidth]{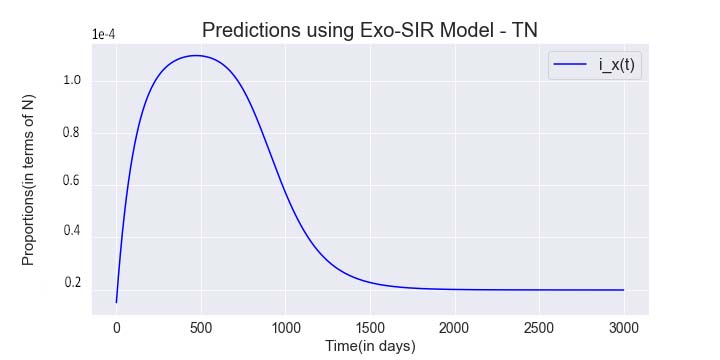}
 	\caption{$i_x$ in Exo-SIR model. Please note that y axis is in the scale of $10^{-4}$.}
 	\label{f108}
\end{subfigure}
\caption{plots for the state of Tamil Nadu. Even for the small values of $i_x$, the peak of the $i_e$ is larger and occurs early.}
\label{fig:fig3}
\end{figure*}

\begin{figure*}
\begin{subfigure}{.33\textwidth}
  \centering
   		\includegraphics[width=\textwidth]{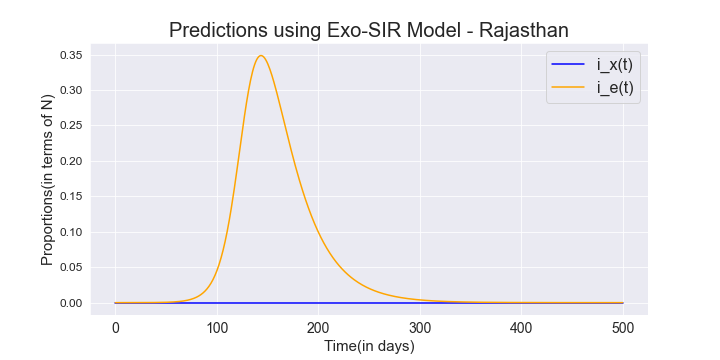}
 	\caption{Exo-SIR model with $i_x$. The values of $i_x$ are very small for the scale of this plot. Hence it is plotted separately. Please refer the Figure~\ref{f105}}
 	\label{f107}
\end{subfigure}%
\begin{subfigure}{.33\textwidth}
  \centering
		\includegraphics[width=\textwidth]{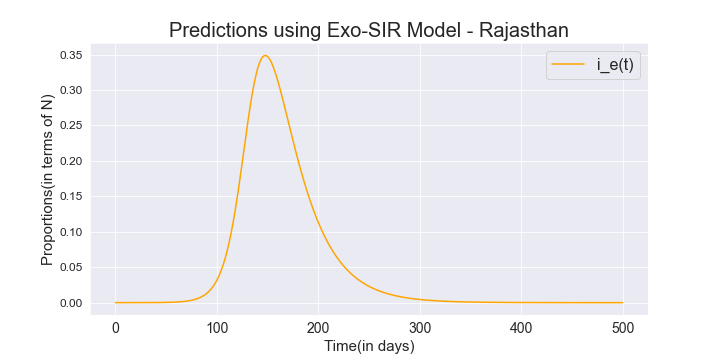}
 	\caption{Exo-SIR model without $i_x$ }
 	\label{f106}
\end{subfigure}
\begin{subfigure}{.33\textwidth}
  \centering
  \includegraphics[width=\textwidth]{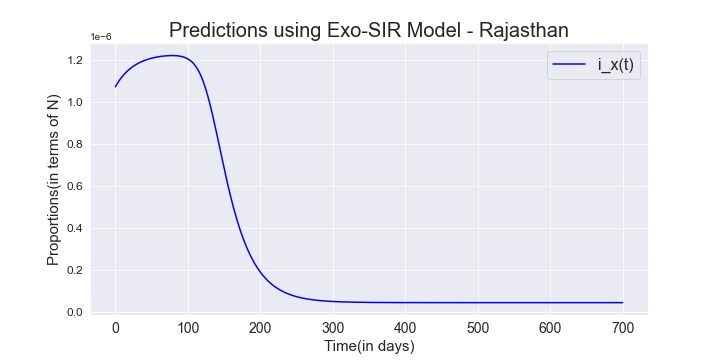}
 	\caption{ $i_x$ in Exo-SIR model. Please note that the y axis is in the scale of $10^{-6}$.}
 	\label{f105}
\end{subfigure}

\caption{plots for the state of Rajasthan. Even for the small values of $i_x$, the peak of the $i_e$ is larger and occurs early.}
\label{f104}
\end{figure*}

\begin{figure*}
\begin{subfigure}{.33\textwidth}
  \centering
   		\includegraphics[width=\textwidth]{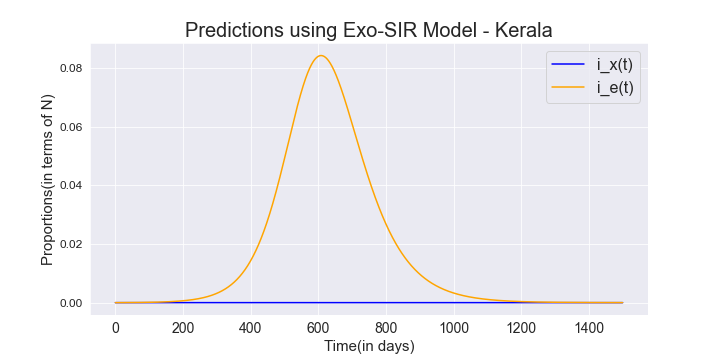}
 	\caption{Exo-SIR model with $i_x$.  The values of $i_x$ are very small for the scale of this plot. Hence it is plotted separately. Please refer the Figure~\ref{f101}}
 	\label{f103}
\end{subfigure}%
\begin{subfigure}{.33\textwidth}
  \centering
		\includegraphics[width=\textwidth]{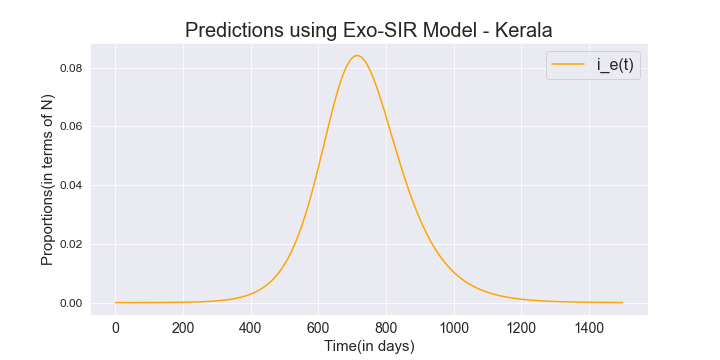}
 	\caption{Exo-SIR model without $i_x$ }
 	\label{f102}
\end{subfigure}
\begin{subfigure}{.33\textwidth}
  \centering
  \includegraphics[width=\textwidth]{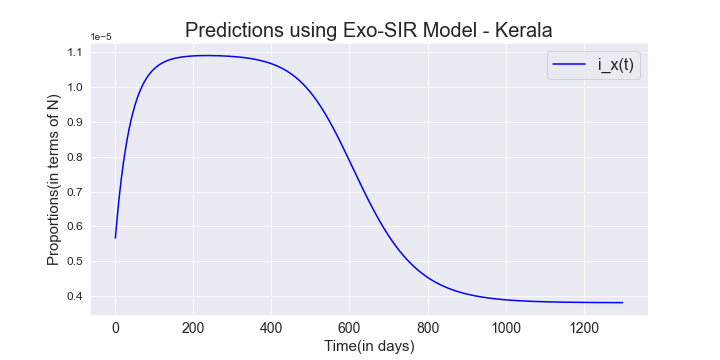}
 	\caption{$i_x$ in Exo-SIR model. Please note that the y axis is in the scale of $10^{-5}$.}
 	\label{f101}
\end{subfigure}

\caption{plots for the state of Kerala. Even for the small values of $i_x$, the peak of the $i_e$ is larger and occurs early.}
\label{fig:fig4}
\end{figure*}

For the state of Tamil Nadu, the Exo-SIR model plots in the presence and absence of $i_x$ are plotted in Figure~\ref{f110} and Figure~\ref{f109} respectively. For the state of Rajasthan, the plots of the Exo-SIR model in the presence and absence of $i_x$ are plotted in Figure~\ref{f107} and Figure~\ref{f106} respectively. For the state of Kerala, the plots of the Exo-SIR model in the presence and absence of $i_x$ are plotted in Figure~\ref{f103} and Figure~\ref{f102}, respectively. In all these plots, we can see that $i_x$ is very small compared to $i_e$. Yet, $i_x$ is having an impact on $i_e$. $I_x$ is plotted separately in Figure~\ref{f101}, Figure~\ref{f105} and Figure~\ref{f108}.

\begin{table}
  \caption{Impact of $I_x$ on $I_e$ in the state of Tamil Nadu}
  \label{t1}
  \begin{tabular}{|l|l|l|}
    \hline
      & peak value & peak tick \\
    \hline
    Exo SIR with $i_x$ & $0.17142523$ & $907$ Days \\
    Exo SIR without $i_x$ & $0.17104498$ & $1351$ Days\\
    \hline
  \end{tabular}
\end{table}


\begin{table}
  \caption{Impact of $I_x$ on $I_e$ in the state of Rajasthan}
  \label{t2}
  \begin{tabular}{|l|l|l|}
    \hline
      & peak value & peak tick \\
    \hline
   
    Exo SIR with $i_x$ & $0.3487077$ & $143$ Days\\
    Exo SIR without $i_x$ & $0.3486663$ & $147$ Days\\
    \hline
  \end{tabular}
 \end{table}

\begin{table}
  \caption{Impact of $I_x$ on $I_e$ in the state of Kerala}
  \label{t3}
  \begin{tabular}{|l|l|l|}
    \hline
      & peak value & peak tick \\
    \hline
   
    Exo SIR with $i_x$ & $0.0842470797$ & $608$ Days\\
    Exo SIR without $i_x$ & $0.0841428705$ & $715$ Days\\
    
    \hline
  \end{tabular}
\end{table}


The peak tick and peak values corresponding to the Exo-SIR model in the presence and absence of $i_x$ for Tamil Nadu, Rajasthan and Kerala are mentioned in Table~\ref{t1}, Table~\ref{t2} and Table~\ref{t3} respectively. In all the tables, we can see that the peak value of $i_e$ is higher when the case of $i_x$ is present. Also, we can see that the peak tick of $i_e$ is earlier for the instance when $i_x$ is present.

\section{Conclusion}

In this evidential study, we introduced the Exo-SIR model. Unlike the SIR model, the Exo-SIR model differentiates between the endogenous and exogenous spread of virus/information. We studied the model in the following ways:

\begin{enumerate}
    \item Theoretical analysis 
    \item Simulation with considering the contact network of the population is a scale free network
    \item Simulation without considering the contact network
    \item Implementation of the Exo-SIR model on real data about the spread of Covid19 in India.
\end{enumerate}

We found that all the four analyses mentioned here converge to the same result: the peak comes early with higher value when the exogenous source is present. This shows that if a Government is preparing to handle a pandemic in the presence of exogenous sources, then Exo-SIR will predict the peak to be earlier and larger compared to the prediction using the SIR model. This information is crucial for the Government because if they follow the SIR model, they will be caught off guard as they will be preparing for a later and smaller peak. Also, we studied the impact of exogenous diffusion on endogenous virus/information diffusion. 

We found that exogenous diffusion impacts endogenous diffusion. If there are exogenous sources of infection like in the case of Covid19, then we should use the Exo-SIR model to estimate the scenario better. This will help the government allocate its resources in a better way as the endogenous and exogenous spread needs different actions to stop them. 


As of now, we have compared our model only with SIR model. As future work, we would like to compare our model with more sophisticated methods like SEIR model and SEYAR model. Also, we would like to study the difference between the asymptomatic and symptomatic infected population. Also, we would like to analyze the impact of a sub-event in India -- the mass movement of migrated laborers across the states in India. 


\section{Acknowledgement}
 Amit Sheth and Manas Gaur are supported by the  National  Science  Foundation  (NSF) award  EAR  1520870:  Hazards  SEES:  Social and  Physical Sensing   Enabled   Decision   Support for   Disaster   Management and  Response.  Any opinions,  findings,  and conclusions/recommendations expressed in this material are those of the author(s)  and do not necessarily reflect the views of the NSF.

\bibliographystyle{ACM-Reference-Format}
\bibliography{main}

\end{document}